\documentclass[
	aip,       
	rsi,       
	reprint,   
	amsmath,   
	amssymb,   
	a4paper,   
    nofootinbib
]{revtex4-2}

\usepackage{dcolumn}
\usepackage{xspace}

\usepackage[T1]{fontenc}
\usepackage[utf8]{inputenc}
\usepackage[english]{babel}

\usepackage{mathtools}
\usepackage{siunitx}

\usepackage{graphicx}

\usepackage{booktabs}

\usepackage{hyperref}
\usepackage{cleveref}

\usepackage{glossaries}
\setacronymstyle{long-short}
\glsdisablehyper

\newcommand{\cm}{\,\si{\cm}}

\newcommand{\ms}{\,\ensuremath{\mathrm{ms}}}
\newcommand{\kHz}{\,\si{\kHz}}
\newcommand{\MHz}{\,\si{\MHz}}
\newcommand{\mHz}{\,\si{\mHz}}

\newcommand{\Hz}{\,\ensuremath{\mathrm{Hz}}}

\newcommand{\s}{\,\ensuremath{\mathrm{s}}}
\newcommand{\ns}{\,\si{\nano\s}}
\newcommand{\K}{\,\si{\K}}

\newcommand{\ohm}{\,\si{\ohm}}

\newcommand{\mohm}{\,\si{\mohm}}

\newcommand{\pF}{\,\si{\pF}}

\newcommand{\degree}{^\circ}
\newcommand{\tesla}{\,\si{\tesla}}
\newcommand{\kelvin}{\,\si{\kelvin}}

\newcommand{\V}{\,\si{\volt}}

\newcommand{\dB}{\,\si{\decibel}}

\newcommand{\minute}{\,\si{\minute}}
\newcommand{\hour}{\,\si{\hour}}

\crefname{equation}{Eq.}{equations}
\Crefname{equation}{Equation}{Equations}
\crefname{table}{Tab.}{Tables}
\Crefname{table}{Table}{Tables}
\crefname{figure}{Fig.}{Figures}
\Crefname{figure}{Figure}{Figures}

\newcommand{\stemlab}{STEMlab\xspace}
\newcommand{\pentatrap}{\textsc{Pentatrap}\xspace}
\newcommand{\alphatrap}{\textsc{Alphatrap}\xspace}
\newcommand{\graphpanel}{graph\xspace}

\newacronym{fpga}{FPGA}{field-programmable gate array}
\newacronym{adc}{ADC}{analog-to-digital converter}
\newacronym{dac}{DAC}{digital-to-analog converter}
\newacronym{fir}{FIR}{finite impulse response}
\newacronym{dds}{DDS}{direct digital synthesis}
\newacronym{snr}{SNR}{signal-to-noise ratio}
\newacronym{apnp}{APnP}{Axial PnP}
\newacronym{pnp}{PnP}{Pulse and Phase}
\newacronym{rf}{RF}{radio frequency}
\newacronym{iq}{IQ}{inphase/quadrature}
\newacronym{fifo}{FIFO}{first-in-first-out}
\newacronym{lo}{LO}{local oscillator}
\newacronym{dma}{DMA}{direct memory access}
\newacronym{zeroIF}{zero IF}{zero intermediate frequency}
\newacronym{vga}{VGA}{variable gain amplifier}
\newacronym{pll}{PLL}{phase-locked loop}
\newacronym{fft}{FFT}{fast Fourier transform}
\newacronym{soc}{SoC}{system on chip}
\newacronym{axi}{AXI}{\textit{Advanced eXtensible Interface}\cite{xilinxUG761AXIReference2012}}
\newacronym{cic}{CIC}{cascaded integrator-comb}
\newacronym{seo}{SEO}{self-excited oscillator}
\newacronym{pwm}{PWM}{pulse-width modulated}

\begin{document}

\title{A Digital Feedback System for Advanced Ion Manipulation Techniques in Penning Traps}

\author{Jost Herkenhoff}
\email{jost.herkenhoff@mpi-hd.mpg.de}
\affiliation{Max-Planck-Institut für Kernphysik, Saupfercheckweg 1, 69117 Heidelberg, Germany}

\author{Menno Door}
\affiliation{Max-Planck-Institut für Kernphysik, Saupfercheckweg 1, 69117 Heidelberg, Germany}

\author{Pavel Filianin}
\affiliation{Max-Planck-Institut für Kernphysik, Saupfercheckweg 1, 69117 Heidelberg, Germany}

\author{Wenjia Huang}
\altaffiliation[Present address: ]{Advanced Energy Science and Technology Guangdong Laboratory, Huizhou 516003, China}
\affiliation{Max-Planck-Institut für Kernphysik, Saupfercheckweg 1, 69117 Heidelberg, Germany}

\author{Kathrin Kromer}
\affiliation{Max-Planck-Institut für Kernphysik, Saupfercheckweg 1, 69117 Heidelberg, Germany}

\author{Daniel Lange}
\affiliation{Max-Planck-Institut für Kernphysik, Saupfercheckweg 1, 69117 Heidelberg, Germany}
\affiliation{Fakultät für Physik und Astronomie, Universität Heidelberg, Im Neuenheimer Feld 226, 69120 Heidelberg, Germany}

\author{Rima X. Schüssler}
\altaffiliation[Present address: ]{Van der Waals-Zeeman Institute, Institute of Physics, University of Amsterdam, 1098 XH Amsterdam, the Netherlands}
\affiliation{Max-Planck-Institut für Kernphysik, Saupfercheckweg 1, 69117 Heidelberg, Germany}

\author{Christoph Schweiger}
\affiliation{Max-Planck-Institut für Kernphysik, Saupfercheckweg 1, 69117 Heidelberg, Germany}

\author{Sergey Eliseev}
\affiliation{Max-Planck-Institut für Kernphysik, Saupfercheckweg 1, 69117 Heidelberg, Germany}

\author{Klaus Blaum}
\affiliation{Max-Planck-Institut für Kernphysik, Saupfercheckweg 1, 69117 Heidelberg, Germany}

\date{\today}

\begin{abstract}
	The possibility to apply active feedback to a single ion in a Penning trap using a fully digital system is demonstrated. 
	Previously realized feedback systems rely on analog circuits that are susceptible to environmental fluctuations and long term drifts, as well as being limited to the specific task they were designed for.
	The presented system is implemented using an FPGA-based platform (\stemlab), offering greater flexibility, higher temporal stability and the possibility for highly dynamic variation of feedback parameters. 
	The system's capabilities were demonstrated by applying feedback to the ion detection system primarily consisting of a resonant circuit. This allowed shifts in its resonance frequency of up to several kHz and free modification of its quality factor within two orders of magnitude, which reduces the temperature of a single ion by a factor of 6. Furthermore, a phase-sensitive detection technique for the axial ion oscillation was implemented, which reduces the current measurement time by two orders of magnitude while simultaneously eliminating model-related systematic uncertainties.
	The use of FPGA technology allowed the implementation of a fully-featured data acquisition system, making it possible to realize feedback techniques that require constant monitoring of the ion signal. This was successfully used to implement a single-ion self-excited oscillator.
	\\
	\\
	\emph{
		The following article has been accepted by Review of Scientific Instruments. After it is published, it will be found at \href{https://aip.scitation.org/journal/rsi}{https://aip.scitation.org/journal/rsi}.}
\end{abstract}

\maketitle


\section{Introduction}

Penning traps are used, among other things, for high-precision atomic mass measurements \cite{brownGeoniumTheoryPhysics1986, blaumHighaccuracyMassSpectrometry2006, myersHighPrecisionAtomicMass2019}, tests of fundamental theories such as quantum electrodynamics \cite{arapoglouFactorBoronlikeArgon2019, volotkaProgressQuantumElectrodynamics2013} and the CPT invariance theorem \cite{smorraPartsperbillionMeasurementAntiproton2017, schneiderDoubletrapMeasurementProton2017, gabrielsePrecisionMassSpectroscopy1999}, as well as for the determination of fundamental constants \cite{blaumPenningTrapsVersatile2009, myersHighPrecisionAtomicMass2019, sturmHighprecisionMeasurementAtomic2014, hannekeNewMeasurementElectron2008, heisseHighPrecisionMeasurementProton2017}. 
Thereby, the frequencies of an ion's motion in the electrostatic and magnetic field of a Penning trap are measured. 
The mass uncertainty mainly arises from the uncertainty of the frequency measurement, which in turn depends on how precisely the ion motion can be manipulated and cooled. To that end, active feedback can be applied, where the signal of the ion oscillation is detected, amplified, phase-shifted and subsequently coupled back onto the ion itself.
This highly versatile technique allows precise control over the oscillation amplitude \cite{dehmeltSelfexcitedMonoionOscillator1986}, enables feedback cooling down to sub-kelvin temperatures \cite{sturmGfactorElectronBound2012}, or can directly manipulate certain parameters like the \gls{snr} or the damping characteristics of the ion detection system \cite{krackeDetectionIndividualSpin2012, rainvillePreciseMeasurementsMasses2001}.
Previously realized feedback systems utilize analog circuits\cite{dehmeltSelfexcitedMonoionOscillator1986, sturmGfactorElectronBound2012, krackeDetectionIndividualSpin2012, rainvillePreciseMeasurementsMasses2001, dursoCoolingSelfexcitationOneelectron2003, dursoFeedbackCoolingOneElectron2003} which are specifically tailored to only one of the feedback applications mentioned above.

In this paper, a novel feedback system based on digital signal processing is presented, offering greater flexibility and extended capabilities compared to conventional analog techniques. High-speed analog-to-digital and digital-to-analog converters (ADCs\glsunset{adc} and DACs\glsunset{dac}) are used for signal conversion, allowing a \gls{fpga} to apply signal transformations like filtering and phase shifting in real-time in the digital domain. A multi-core processor, which is directly interfaced to the \gls{fpga} fabric, is used for signal analysis and can be utilized to implement adaptive filters or to realize complex measurement cycles without the need for auxiliary devices.
The digital architecture pushes the boundaries of currently possible ion manipulation techniques and opens up the possibility for improved frequency measurement techniques based on direct phase determination of the ion oscillation.

The presented feedback system is deployed at the Penning-trap experiment \pentatrap\cite{rouxTrapDesignPENTATRAP2011}, located at the Max Planck Institute for Nuclear Physics, where it will contribute to achieving the aimed precision of a few parts in $10^{12}$ for mass-ratio measurements of highly charged ions in the medium-heavy to heavy mass range.
\pentatrap provides atomic masses to test, among others, quantum electrodynamics in the regime of extreme electric fields \cite{reppPENTATRAPNovelCryogenic2012} and contributes to the determination of an upper limit for the neutrino mass \cite{blaumPenningTrapsVersatile2009, eliseevResonantEnhancementNeutrinoless2011, gastaldoElectronCapture163Ho2017}.

\section{Penning Traps \& Active Feedback}
\label{sec:penningtraps-and-feedback}
\begin{figure}
	\includegraphics{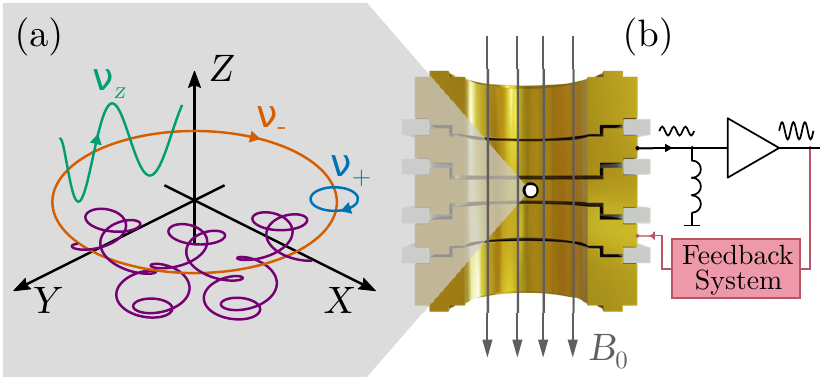}
	\caption{(a) Trajectory of the ion motion (magenta), consisting of the axial (green), cyclotron (blue) and magnetron oscillation (orange). Figure (b) shows a cut through a sketched Penning-trap electrode structure with a schematic of the attached detection system for the axial motion. The red signal path visualizes the feedback system in its configuration to apply direct feedback to the ion.}
	\label{fig:penning-trap}
\end{figure}

Penning traps confine a single ion using the superposition of a homogeneous magnetic field with field strength $B_0$ and an electrostatic quadrupole field created by a stack of electrodes as shown in Fig. \hyperref[fig:penning-trap]{\ref*{fig:penning-trap}b}. The ion follows a trajectory consisting of three independent harmonic eigenmotions, namely the axial, the modified cyclotron and the magnetron motion, depicted in Fig. \hyperref[fig:penning-trap]{\ref*{fig:penning-trap}a}. By measuring their respective frequencies $\nu_z$, $\nu_+$ and $\nu_-$, the mass $m$ of the ion with charge $q$ can be calculated using the relation \cite{brownGeoniumTheoryPhysics1986}
\begin{equation}
\sqrt{\nu_z^2 + \nu_+^2 + \nu_-^2} = \frac{q B_0}{2 \pi \, m} = \nu_c\,,
\end{equation}
with $\nu_c$ being the free-space cyclotron frequency.

\begin{figure}
	\includegraphics{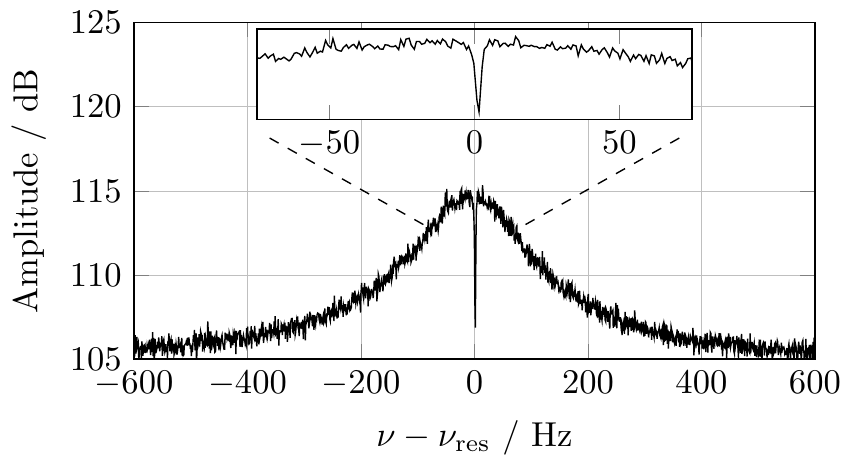}
	\caption{Measured frequency spectrum of the detection system output signal. The underlying ``bell shaped'' curve originates from the  thermal noise of the resonator, shaped by its frequency dependent impedance. A dip signature is visible at $\nu \approx \nu_\text{res} \approx 740\,\kHz$, which is the result of an ion in thermal equilibrium with the detection system.}
	\label{fig:dip-theory}
\end{figure}

The individual oscillatory motions induce image currents on the order of a few femtoampere between the trap electrodes\cite{shockleyCurrentsConductorsInduced1938, ramoCurrentsInducedElectron1939}. These small currents can be converted into a measurable voltage by connecting a high impedance $Z$ to an electrode that is offset from the ion's center of motion in the direction of oscillation to be detected\cite{marshallFourierTransformIon1998, winelandPrinciplesStoredIon1975}. 
At the \pentatrap experiment, only the axial motion is detected with this technique by using a pickup electrode that is axially offset.
The impedance is realized using a superconducting coil $L$, which forms a resonant LCR circuit (hereinafter called \emph{resonator}) together with the parasitic capacitance $C$ of the detection electrode. Residual ohmic and dielectric losses lead to a finite resistance $R$, yielding a quality factor $Q=R/\sqrt{C/ L}$ of about 4400.
The resonator's frequency-dependent impedance is then given by
\begin{equation}
\label{eq:impedance}
Z(\nu) = R \left[ 1 + i Q \left(\frac{\nu}{\nu_\text{res}} - \frac{\nu_\text{res}}{\nu}\right)\right]^{-1}\,,
\end{equation}
with $\nu_\text{res} = (2 \pi \sqrt{LC})^{-1}$ being its resonance frequency.
Tuning the ion's axial frequency into resonance results in a purely resistive impedance $Z(\nu_\text{res}) = R$. The voltage drop across the resonator is detected by a cryogenic low-noise amplifier, making the axial oscillation visible as a peak at $\nu_z$ in the frequency spectrum of the amplified signal.
The current induced in the trap electrode causes energy from the ion's axial motion to get dissipated in the detection system, leading to an exponential decrease of its amplitude with a time constant $\tau$ that is proportional to $1 / \text{Re}\{Z\}$. Due to the associated reduction of accessible phase space volume, this process is commonly referred to as \emph{resistive cooling} \cite{winelandPrinciplesStoredIon1975, itanoCoolingMethodsIon1995}. It is counteracted by the incoherent thermal noise of the detection system, leading to a finite equilibrium temperature of the ion that is equal to the temperature of the detection system. As the \pentatrap experiment is cooled by thermal contact to a liquid helium bath at atmospheric pressure, the axial mode is expected to equilibrate at $T_z = 4.2\K$\footnote{Ignoring any further internal and external noise sources, which would increase the equilibrium temperature of the ion.}. At this time, the ion is no longer visible as a peak in the frequency spectrum.
However, it can still be detected, since the thermalized ion effectively shortens the thermal noise of the resonator, leading to a dip in its spectrum,  shown in \cref{fig:dip-theory}. 
While in theory the noise amplitude at the dip center is expected to be zero, in reality it is elevated by the amplifier noise.
The \gls{snr} of this dip-detection technique is defined as the ratio  $n_\text{res} / n_\text{dip}$ of the resonator lineshape noise density $n_\text{res}$ at $\nu_z$ and the locally minimum noise density of the ion dip $n_\text{dip}$.

In order to measure the cyclotron and magnetron frequencies, they are independently coupled to the axial mode using \emph{sideband coupling}\cite{brownGeoniumTheoryPhysics1986}.
The coupled modes experience resistive cooling from the interaction with the resonator and equilibrate at temperatures of\cite{djekicTemperatureMeasurementSingle2004}

\begin{equation}
\label{eq:radial-temperatures}
T_+ \simeq \frac{\nu_+}{\nu_z} T_z \quad \text{and} \quad T_- \simeq -\frac{\nu_-}{\nu_z} T_z\,.
\end{equation}
The negative sign of the magnetron temperature $T_-$ arises from the inverted energy level scheme of the magnetron motion\cite{brownGeoniumTheoryPhysics1986}.

While the signal from the axial detection circuit is primarily used for frequency determination, it can simultaneously be used to apply active feedback to the ion.
For this purpose, the detected ion signal is phase shifted, amplified or attenuated and subsequently fed back to an electrode of the Penning trap. 
This produces an additional ``ion derived'' excitation field within the trap that directly influences the ion motion. While other configurations are possible, the following discussion assumes the feedback to be applied to an electrode that is opposite to the detection electrode with respect to the ion center position, as depicted in \cref{fig:penning-trap}.
The axial motion of an ion under active feedback is then described by the equation
\begin{equation}
\ddot{z} + \gamma \left(1 - G\right) \dot{z} + (2 \pi \nu_z)^2 \, z = 0\,,
\end{equation}
with $\gamma \sim \frac{1}{\tau}$ being the damping due to resistive cooling.
The active feedback loop itself is represented by a generalized complex feedback gain $G = |G| e^{i\phi}$, with $\phi$ being the introduced phase shift and $|G|$ the gain magnitude accounting for all signal amplifications or attenuations within the ion detection system, the feedback loop itself, as well as the coupling strength between the ion and the trap electrodes. It can be seen that the application of feedback with a phase of $0^\circ$ or $180^\circ$ can be used to arbitrarily modify the damping of the ion:
\begin{equation}
\label{eq:ion-damping-feedback}
\gamma^\text{fb} = \gamma \left( 1 - \text{Re} \{G\}\right)\,.
\end{equation}
This opens up the possibility to achieve zero or even negative damping, resulting in a self-excited oscillation. The fluctuation-dissipation theorem \cite{callenIrreversibilityGeneralizedNoise1951, kuboFluctuationdissipationTheorem1966} states, that the damping of the ion is inherently connected to its temperature, revealing that feedback with $0 < |G| < 1$ and $\phi = 0^\circ$ is able to reduce the axial temperature of the ion below its equilibrium temperature without feedback \cite{dursoFeedbackCoolingOneElectron2003}:
\begin{equation}
T_z^\text{fb} = T_z\,\left(1- \text{Re}\{G\}\right)\,.
\end{equation}

The application of $180^\circ$ feedback increases the damping, resulting in a faster cooling rate at the expense of a higher equilibrium temperature.
Phases other than $0^\circ$ and $180^\circ$ lead to a detuning of the axial frequency $\nu_z$, which is often undesirable. Thus, a fine grained adjustable phase shift with high temporal stability is required in order to precisely achieve $0^\circ$ or $180^\circ$ feedback.

Care must be taken, as the feedback signal fed into the Penning trap can couple directly back to the resonator through parasitic inter-electrode capacitances, leading to modifications of its properties.
This can be avoided using a secondary feedback path fed into another trap electrode or coupled directly to the resonator using a discrete capacitor. By carefully tuning the phase and gain relations between both feedback paths so that they destructively interfere at the detection electrode, the feedback signal as seen by the detection system can be canceled out almost completely while maintaining the desired effect on the ion \cite{dursoFeedbackCoolingOneElectron2003}.

\begin{figure}
	\includegraphics{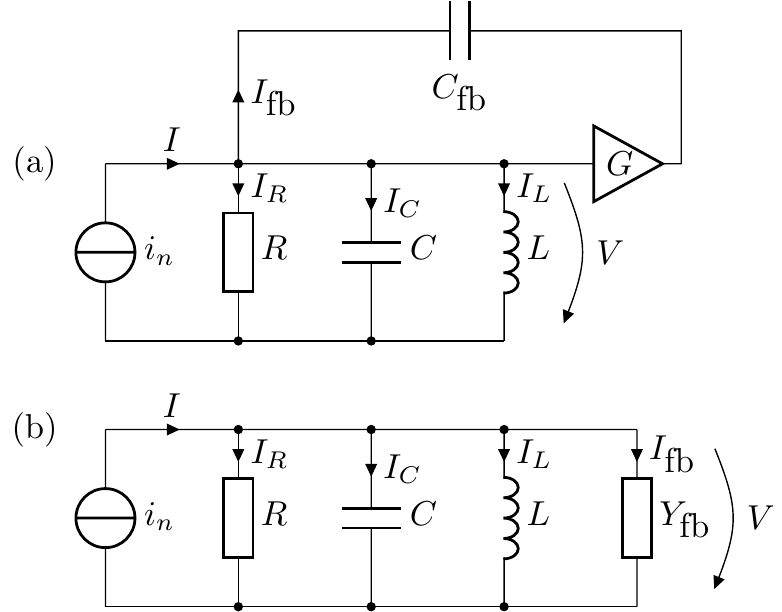}
	\caption{Schematic diagram of the resonator (consisting of $R$, $C$ and $L$) used in the ion detection system with applied active feedback (a) and its equivalent representation modeling the feedback loop with a complex admittance $Y_\text{fb}$ (b). The current source $i_n$ represents the thermal noise of $R$. For details see text.}
	\label{fig:resonator-feedback}
\end{figure}

On the other hand, the previously mentioned effect of active feedback coupling back to the detection system can be deliberately utilized to modify its characteristics.
For that, a discrete capacitor $C_\text{fb}$ is used to achieve direct coupling to the resonator.
Figure \hyperref[fig:resonator-feedback]{\ref*{fig:resonator-feedback}a} depicts the schematic of the resonator with the feedback loop, which can be equivalently modeled as an entirely passive circuit by introducing the complex admittance $Y_\text{fb} = i\omega C_\text{fb} (1-G)$, as shown in Fig. \hyperref[fig:resonator-feedback]{\ref*{fig:resonator-feedback}b}.
Using this model, an effective resistance
\begin{equation}
R_\text{eff} = \frac{R}{1 + 2\pi\nu C_\text{fb} R \, \text{Im}\{G\}}
\end{equation}
can be defined, showing that the resistance of the resonator can be arbitrarily modified by changing the gain $G$. When applying feedback such that $R_\text{eff} < R$, the Johnson-Nyquist noise \cite{johnsonThermalAgitationElectricity1928, nyquistThermalAgitationElectric1928} of the circuit gets actively suppressed, resulting in a detection system that ``looks'' colder to the ion than its actual physical temperature $T_0$:
\begin{equation}
\label{eq:resonator-temperature}
T_\text{eff} = T_0 \, \frac{R_\text{eff}}{R}\,.
\end{equation}
The minimum achievable temperature is limited by electrical noise within the feedback path, dominantly originating from the cryogenic amplifier, which is not included in \cref{eq:resonator-temperature}. For a detailed discussion see \onlinecite{dursoFeedbackCoolingOneElectron2003}.

Feedback with $\phi = 0^\circ$ or $180^\circ$ effectively alters the capacitance of the resonator, resulting in a varying resonance frequency
\begin{equation}
\label{eq:resonator-frequency}
\nu_\text{res,fb} = \frac{\nu_\text{res}}{\sqrt{1 + C / C_\text{fb} \left( 1 - \text{Re} \{G\}\right)}}\,,
\end{equation}
with $\nu_\text{res}$ being the resonance frequency without feedback. 
The ion can therefore be completely decoupled from the detection system by introducing a large resonance frequency detuning.

\section{Implementation of the digital feedback system}

\subsection{Hardware}

The digital feedback system is implemented using the \stemlab 125-14 platform\cite{redpitayateamRedPitayaWebsite2021} (formerly: Red Pitaya) for analog/digital conversion and signal processing. It was already successfully used in various real-time applications in physics, including laser and frequency comb stabilization\cite{tourigny-planteOpenFlexibleDigital2018, shawVersatileDigitalApproach2019} and lock in amplifiers\cite{stimpsonOpensourceHighfrequencyLockin2019}, but to our knowledge has never been used for the direct manipulation of single ions.
The \stemlab is built around a Zynq 7000\cite{xilinxZynq7000SoCData2018} \gls{soc}, combining a dual core ARM A9 CPU with \gls{fpga} fabric on the same device. Additionally, it provides two \glspl{adc} and \glspl{dac} (LTC2145CUP-14 and AD9767) with a resolution of 14\,bit and a sampling rate of $125\MHz$, which are used for the real-time data conversion. Due to the large resulting data rate of $3.5\,\text{Gbit}/\text{s}$, \glspl{fpga} are the platform of choice for the digital feedback system.

\begin{figure}
	\includegraphics{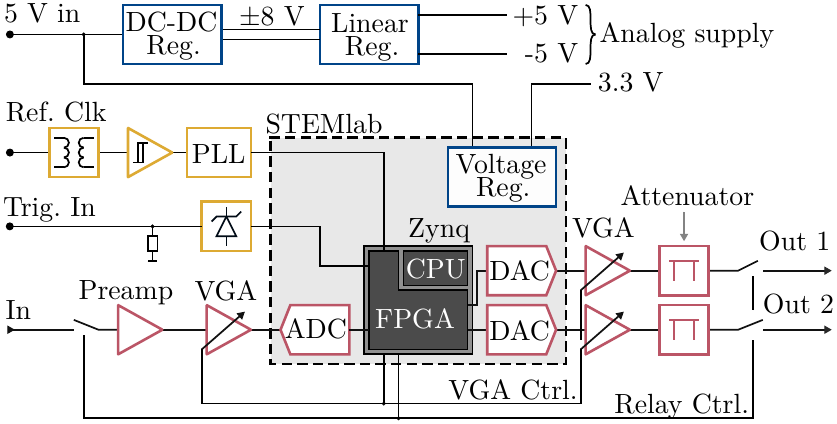}
	\caption{Block diagram of the feedback system's hardware. Red blocks: Feedback path; Yellow blocks: Reference clock and trigger; Blue blocks: Power supply. For details see text.}
	\label{fig:hardware-block-diagram}
\end{figure}

\Cref{fig:hardware-block-diagram} shows the block diagram of the custom hardware built around the \stemlab platform.
As the ion detection system outputs a signal with an amplitude of only a few millivolt, further amplification is necessary to fully utilize the $\pm 1\V$ dynamic range of the \gls{adc}. This is accomplished using an AD8429 instrumentation amplifier with $20\dB$ gain, followed by an AD8336 \gls{vga} with a range of $-10\dB$ to $40\dB$. The \gls{vga} gain is controlled by an analog control voltage generated by a filtered \gls{pwm} output of the \gls{fpga}. This allows one to remotely adjust the input amplitude or enables the implementation of automated gain control.
After the ion signal has been processed by the \gls{fpga} (see \cref{sec:fpga}), the two output datastreams are converted back into the analog domain using two \glspl{dac}. The \stemlab provides a DC offset removal circuit, to shift the \glspl{dac} unipolar output range to a symmetric output of $\pm 1\V$. However, this circuit is reported to couple excess noise into the analog outputs, which is why we removed it as described in Ref. \onlinecite{userlneuhausRedPitayaDAC2021}. Instead, we used a serial AC coupling capacitor to symmetrize the output voltage. As a result, its frequency response is cutoff below $\approx 100\Hz$, which is sufficiently far away from the frequency range of interest of a few $100\kHz$.
It is advisable to utilize the full output voltage span of the \gls{dac} in order to make best use of its available resolution and therefore achieve the highest possible SNR. For adapting the output signal to different amplitude level requirements, each analog output is connected to two independently controllable AD8336 \glspl{vga} with a range of $-26\dB$ to $34\dB$. Additionally, a set of fixed attenuators ranging from 2\,dB to 16\,dB are available, which can be manually engaged into the output paths to accommodate an even larger output amplitude range.
The analog circuitry can be electrically disconnected from its input and output connectors using digitally controlled relays. When the feedback system is not in use, this can be helpful to eliminate potential noise sources, which are often a major concern in high-sensitivity experiments like Penning traps. However, the presented system was not found to introduce significant excess noise into the experimental setup.

In its factory configuration, the \stemlab uses an onboard $125\MHz$ crystal oscillator to clock the \gls{fpga}, \glspl{adc} and \glspl{dac}. Since temporal stability is of great importance for the feedback system, additional circuitry was designed to allow the \stemlab to derive its timing from an external $10\MHz$ rubidium frequency standard. The external clock is galvanically isolated by a transformer, eliminating the risk of ground loops. A clock conditioning circuit, built around an LTC6957 clock buffer, generates a clean rectangular signal that is fed into an AD9552 \gls{pll}, which is used to produce the $125\MHz$ main clock.
The high accuracy of the external clock source allows the feedback system to also be used for high-precision frequency measurements.
An external trigger signal is connected to a digital input of the \gls{fpga} to allow the feedback system to be synchronized with the measurement sequence of \pentatrap (see section \cref{sec:fpga}). The trigger input is $50\ohm$ terminated and equipped with clamping diodes to protect the \gls{fpga} input from overvoltages.

\begin{figure}
	\includegraphics{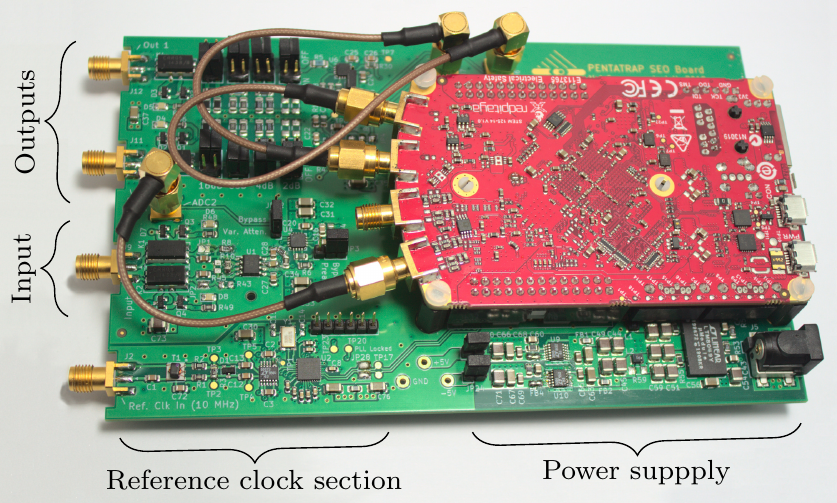}
	\caption{Picture of the assembled digital feedback system. The \stemlab (red PCB) is mounted onto the custom designed PCB using pin headers. The analog input and outputs are connected to the main PCB using SMA cables. The whole assembly is mounted inside an aluminium housing (not shown)}
	\label{fig:pcb}
\end{figure}

The feedback system is powered using an external 5V supply. The input and output amplifiers require a bipolar low-noise supply with $\pm 5\V$. For that, a dual channel DC-DC switchmode converter (LTM8049) generates two intermediate voltages with $\pm 8\V$ out of the external unipolar $5\V$ supply. The switchmode frequency is set to $1.7\MHz$, which is well above the frequency band of interest in order to prevent interference. After filtering with ferrite beads, the intermediate voltage rails are fed into a pair of ultra low-noise linear voltage regulators (LT3045 and LT3094) to generate the final $\pm 5\V$ analog supply.
The reference clock circuitry is powered by a $3.3\V$ supply rail that is generated by an internal voltage regulator of the \stemlab.
A picture of the feedback system is shown in \cref{fig:pcb}.

\subsection{FPGA}
\label{sec:fpga}

The \stemlab's \glspl{adc} and \glspl{dac} are connected to the \gls{fpga} using a parallel interface, offering a low latency datapath. The total propagation delay of the feedback path from pickup electrode to feedback electrode, including the signal processing algorithm, must be minimized, as any introduced latency causes a frequency dependent phase response that directly limits the achievable feedback bandwidth. Thus, the whole \gls{fpga} implementation is specifically tailored to minimize the internal latency.
\begin{figure*}
	\includegraphics{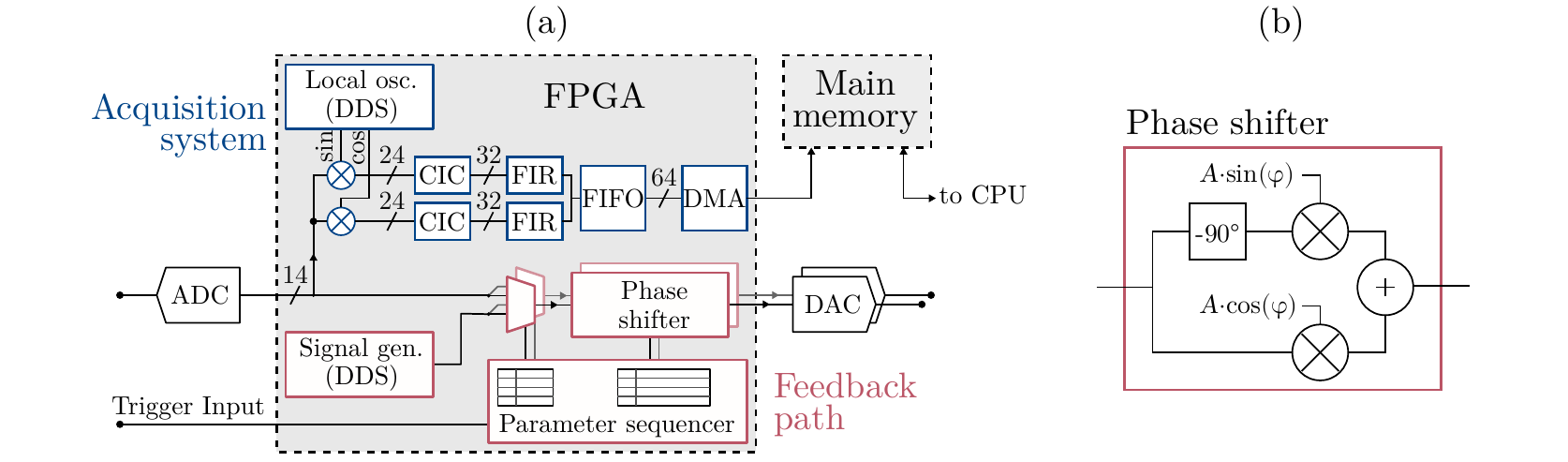}
	\caption{(a) Simplified block diagram of the \gls{fpga} implementation. The blue and red outlined blocks make up the acquisition system and the feedback paths, respectively. Some datapaths are marked with a diagonal line, with the associated number denoting the word width in bits. (b) Detailed block diagram of the phase shifter implementation.}
	\label{fig:fpga}
\end{figure*}

Figure \hyperref[fig:fpga]{\ref*{fig:fpga}a} shows a block diagram of the \gls{fpga} implementation.
One of the core components of the feedback system are the variable phase shifters.
They are implemented using an \gls{iq} architecture, depicted in Fig. \hyperref[fig:fpga]{\ref*{fig:fpga}b}. A Hilbert transform is used to produce a constant $-90\degree$ phase shifted copy of the input signal, called the quadrature (Q) component. Together with the original signal, called the inphase (I) component, an output signal with arbitrary phase shift can be obtained by summing the inphase and quadrature components with weights of $\cos(\varphi)$ and $\sin(\varphi)$, respectively.

An ideal Hilbert transform has a perfectly flat phase response but cannot be realized due to its non-casual impulse response.
However, for a narrow bandwidth around the frequency of interest, i.e.\ the axial frequency $\nu_z$, it can be approximated by a time delay of $t_{90^\circ} = 90^\circ/360^\circ / \nu_z$.
For other frequencies than $\nu_z$, the desired phase shift deviates from $-90^\circ$ with a slope of $\text d \varphi / \text d \nu = -t_{90^\circ} \cdot 360^\circ$. For an axial frequency of $\nu_z = 740\kHz$, the slope is $\approx 0.12^\circ / \kHz$, which is acceptable considering the narrow bandwidth of the detection system of a few $100\Hz$.
The time delay $t_{90^\circ}$ translates into a delay of $D = \nu_\text{clk} \cdot t_{90^\circ}$ clock cycles, with $\nu_\text{clk} = 125\MHz$ being the \gls{fpga} clock frequency. Since $D$ is generally a non integer number and there are no fractional clock cycles, a linear interpolation filter had to be used to obtain the required sub-clock cycle time resolution.
For that, the input signal is split into two paths that are delayed by $\lfloor D \rfloor$ and $\lceil D \rceil$\footnote{The $\lfloor \cdot \rfloor$ and $\lceil \cdot \rceil$ operators represent the floor and ceiling functions, returning the nearest smaller or greater integer number, respectively.} clock cycles and are subsequently summed together with a weight of $(1-\delta)$ and $\delta$, respectively, with $\delta = D - \lfloor D \rfloor$ being the fractional part of $D$. This structure provides a phase shift of precisely $-90\degree$ for the frequency $\nu_z$, but also forms a low pass filter with a worst case ($\delta = 0.5$) cutoff frequency of $\approx 32\MHz$, which is fortunately way above the axial frequency. 
The fractional delay line is implemented using a \gls{fifo} buffer with two taps at the $\lfloor D\rfloor$th and $\lceil D \rceil$th memory cell, where $D$ can be configured via software.

\begin{figure}
	\includegraphics{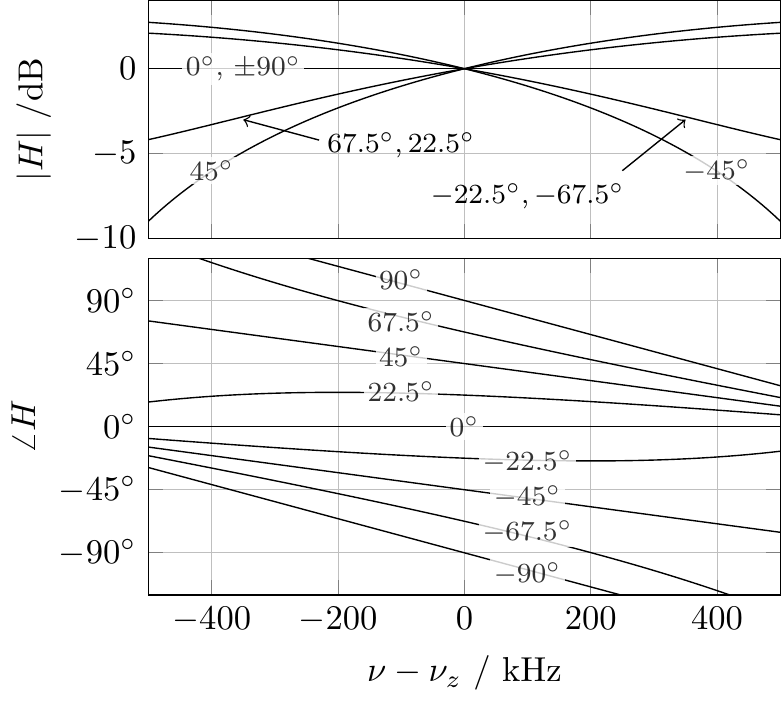}
	\caption{Frequency response of the variable phase shifter for nine different phase shift settings from $-90\degree$ to $90\degree$. The upper \graphpanel shows the magnitude, the lower \graphpanel the phase of the response function $H$. Phase settings outside this range have an identical relative shape of the frequency response except for a constant phase offset, which is why they were omitted for clarity. The fractional delay line was configured for an ion frequency of $\nu_z = 740\kHz$ ($\lfloor D \rfloor = 42$, $\delta = 0.23$). At this frequency, the magnitude $|H|$ is exactly $0\dB$ and the phase $\angle H$ coincidences precisely with the respective configured phase.}
	\label{fig:iq-shifter-freq-response}
\end{figure}

The quadrature component, i.e.\ the output of the fractional delay line, and the inphase component are then weighted by $A\sin(\varphi)$ and $A\cos(\varphi)$ using hardware multipliers. The common scaling factor $A \in [0, 1]$ allows the application of digital attenuation to the output signal.
A complete description of the \gls{iq} phase shifter implementation is given by its impulse response
\begin{equation}
h[n] = A\cdot[\,\underbrace{\cos(\varphi)}_{n=0}, ~ 0,\,\dots\,,~ \underbrace{(1-\delta)\sin(\varphi)}_{n=\lfloor D\rfloor}, ~\underbrace{\delta \sin(\varphi)}_{n=\lceil D\rceil} \, ]\,.
\end{equation}
Evaluating its discrete Fourier transform $H(\nu) \bullet \!\!-\!\!\circ h[n]$ yields the frequency response shown in \cref{fig:iq-shifter-freq-response}. While it exactly matches an ideal phase shifter at $\nu_z$, its response diverges for other frequencies. Using a linear expansion around $\nu_z$, a worst case magnitude slope of $\pm 0.01\dB/\kHz$ for $\varphi_\text{set} = \pm 45\degree$ was found. The worst case phase slope of $\pm 0.12\degree/\kHz$ for $\varphi_\text{set} = \pm 90\degree$ originates from the fractional delay line, as described earlier. For the narrow bandwidth of the ion detection system, these effects can be neglected.

The parameters $A$ and $\varphi$ can be dynamically varied for both phase shifters independently using a parameter sequencer module programmed into the \gls{fpga}. It contains a list of eight software-configurable parameter sets, that are sequentially loaded into the phase shifters with precise timing. The step condition can either be a fixed time delay programmed into the corresponding parameter set or a trigger condition, i.e.\ rising or falling edge of the external trigger input. As the trigger signal assertion is not necessarily synchronized with the \gls{fpga} clock frequency, a worst case timing jitter of one clock cycle ($8\,\ns$) might occur. This effect is eliminated by synchronizing the trigger signal generation to the $10\MHz$ reference clock input.

The parameter sequencer also controls the input source of the feedback paths using a connected multiplexer. It selects either between the \gls{adc} input signal or one of two sinusoidal signals generated by two built-in \gls{dds} modules whose frequency can be configured via software. This can be used to generate sinusoidal pulses for ion excitation, as used in various measurement schemes.

In addition to the feedback paths, the \gls{fpga} contains an acquisition system, which can be used for ion detection.
To efficiently extract the narrow ion signal from the wideband \gls{adc} signal, it is mixed down to a complex signal with \gls{zeroIF} using a digital \gls{iq} mixer architecture. Two hardware multipliers are used to produce the downmixed \gls{iq} signals by mixing the input signal with two, $90\degree$ out of phase, local oscillator signals, generated using a \gls{dds} module with software-configurable frequency.
After downmixing, \gls{cic} filters decimate the data rate, and thus the bandwidth, by a variable factor of 64 to 4096.
To accommodate for the bit growth during downconversion, the signal paths after the \gls{cic} filters are chosen to have a width of 32\,bits.
The \gls{cic} filters have a non-flat passband, which is compensated for using \gls{fir} filters with 309 taps, designed to eliminate the passband droop. The inphase and quadrature paths are then combined into a 64\,bit datastream, which is fed into a buffering \gls{fifo} memory with 2048 entries.
Finally, a \gls{dma} engine writes the buffered datastream into a dedicated region of the main memory, from which the data can be read and further processed in software on the ARM A9 CPUs.

The acquisition system can be triggered in precise synchronization to the feedback sequencer, allowing the implementation of complex measurement cycles, turning the digital feedback system into a stand-alone measuring device.

\subsection{Software}
\label{sec:software}

Due to the tight integration of \gls{fpga} and CPU, the advantages of both can be utilized: While the \gls{fpga} handles the specialized, real-time data processing, the CPU carries out control tasks and augments the data processing with its ability to implement general purpose algorithms.
At its base, the \stemlab's dual core ARM CPU runs the GNU/Linux operating system. On top of that, a custom python library acts as a ``driver'' to provide easy access to the functionality of the feedback system. It can read/write all parameters of the \gls{fpga} modules, which are connected to the CPU memory space by an \gls{axi} bus.
A jupyter notebook\cite{projectjupyterProjectJupyter2021} server can be accessed using the \stemlab's network connection and provides a convenient way to implement measurement scripts.
Additionally, an EPICS\cite{epicsteamEPICSWebsite2020} (Experimental Physics and Industrial Control System) server allows remote control of the feedback parameters and integration into the existing \pentatrap control system.

A standalone software was developed to realize the first digitally controlled self-excited oscillating ion:
In order for the ion to maintain its oscillation amplitude, the feedback gain must be constantly adjusted. For this, the ion signal is sampled and transferred into memory using the built-in acquisition system. The software then calculates its frequency domain spectrum using a \gls{fft} and determines the ion amplitude using a peak detection algorithm. The amplitude deviation from a given setpoint is then fed into a PID control loop to calculate the needed feedback gain. After updating the feedback parameters accordingly, the cycle is repeated. The results of this \gls{seo} implementation are presented in \cref{sec:seo}.

\section{Applications}


\subsection{Resonator modification}

\begin{figure*}
	\includegraphics{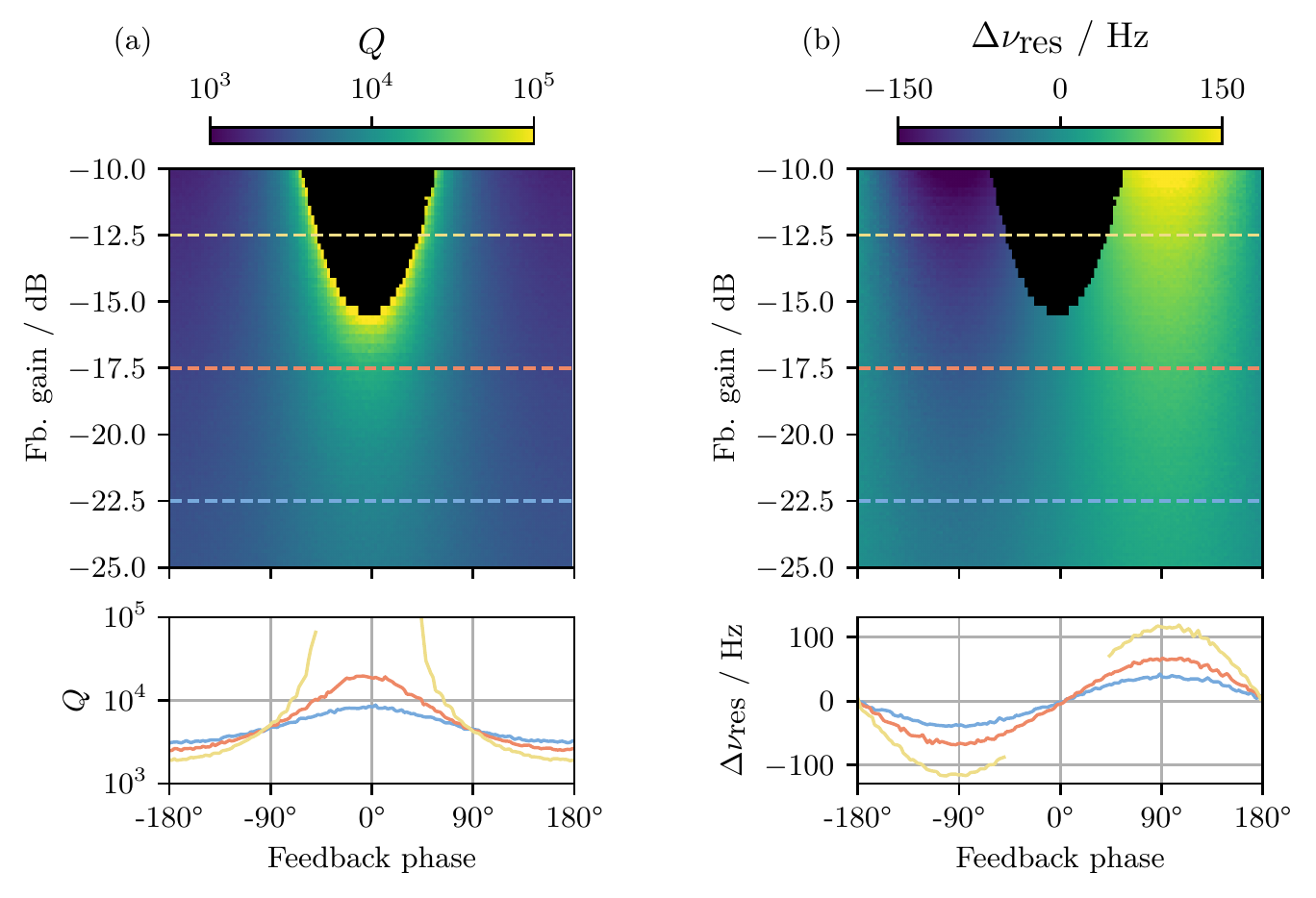}
	\caption{Results of the resonator modification parameter scan, showing the $Q$ value (a) and the detuning of the resonance frequency (b). The heatmap consists of $128\times 128$ pixels, each representing a different feedback gain/phase setting. The black areas indicate the instability regions where the resonator was driven into self-oscillation. The lower \graphpanel{s} show 1D cuts through the heatmap, at the locations indicated by the dashed lines.}
	\label{fig:resonator-mod-heatmap}
\end{figure*}

As described in \cref{sec:penningtraps-and-feedback}, active feedback can be utilized to modify the parameters of the resonator used in the detection system.
To characterize this modification, a parameter space consisting of $128 \times 128$ feedback gain/phase pairs, ranging from -26\dB\ to -10\dB\ and $-180\degree$ to $180\degree$, was scanned, while monitoring the $Q$ value and $\nu_\text{res}$ of the resonator. The feedback signal was connected to a trap electrode, resulting in coupling to the resonator by means of the parasitic capacitance to the pickup electrode.
For each point in the parameter space, a set of 40 spectra with a span of $15\kHz$ centered around the resonator's center frequency and a resolution of $3.5\Hz$ were acquired using the feedback system's built-in acquisition system. Averaging the 40 spectra of each datapoint results in a set of $128 \times 128$ spectra with low noise, which are individually fitted to the mathematical model of the resonator's frequency response. \Cref{fig:resonator-mod-heatmap} shows the obtained results in a 2D heatmap representation.

Compared to the nominal $Q$ value of about 4400, the feedback system can significantly increase it up to $10^5$, as can be seen in  Fig. \hyperref[fig:resonator-mod-heatmap]{\ref*{fig:resonator-mod-heatmap}a}. With such a high Q factor, the damping of the resonator is close to zero, so that it can easily be driven into free oscillations, represented by the black regions in \cref{fig:resonator-mod-heatmap}, potentially causing ion loss.
It must be noted that the arbitrarily increased $Q$ value simultaneously heats up the ion according to \cref{eq:resonator-temperature}, which can lead to undesired systematic errors. However, the high \gls{snr} enables faster measurements, which is advantageous e.g. for ion loading and preparation, where fast user feedback is more important than absolute frequency accuracy.

\begin{figure}
	\includegraphics{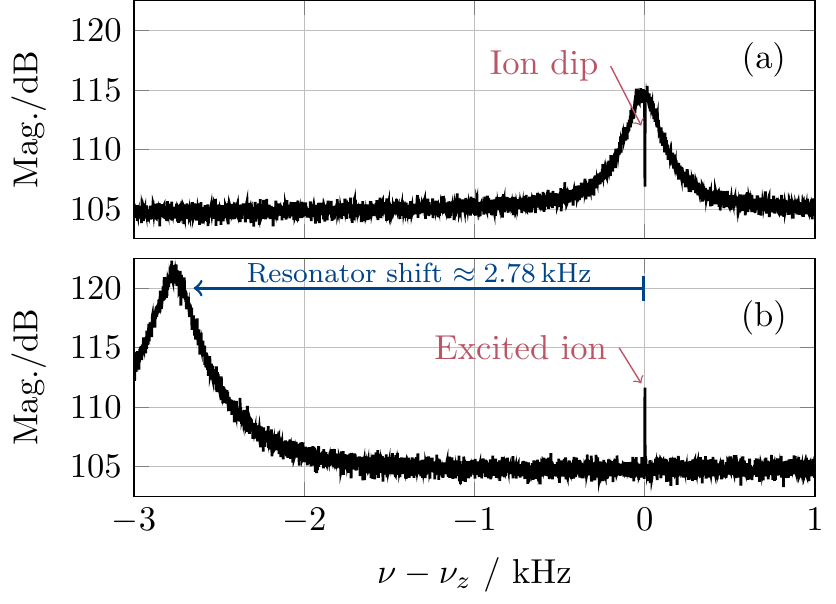}
	\caption{Frequency spectrum of the axial detection system. Plot (a) shows the resonator lineshape without active feedback with a visible dip feature at $\nu_z$ originating from a thermalized ion. Plot (b) shows the resonator lineshape with enabled active feedback, configured for a resonance frequency shift of $\approx 2.78\kHz$. For the presented measurement, the ion was excited to a large amplitude, resulting in an almost infinitely persisting peak signal at $\nu_z$. The resonator-detuning largely inhibits resistive cooling of the ion.}
	\label{fig:resonator-shift-ion}
\end{figure}

Figure \hyperref[fig:resonator-mod-heatmap]{\ref*{fig:resonator-mod-heatmap}b} shows the detuning $\Delta \nu_\text{res}$ of the resonator for different feedback parameters. The results perfectly agree with \cref{eq:resonator-frequency}.
An extreme case of this frequency detuning is shown in \cref{fig:resonator-shift-ion}, where the resonator is shifted by several linewidths. This way the ion can be decoupled from the resonator.

\subsection{Feedback Cooling}

To measure the effect of active feedback cooling, the temperature of the axial motion of a single $^{40}\text{Ar}^{13+}$ ion was measured with different feedback parameters.
The measurements have been carried out at the \alphatrap experiment \cite{sturmALPHATRAPExperiment2019}, as it enables strong feedback coupling to the resonator via dedicated, capacitive feedback paths. Additionally, the availability of a magnetic field inhomogeneity simplifies axial temperature measurements, as described below.

In a macroscopic system, temperature is defined as the distribution of kinetic energy across an ensemble of particles. The ergodic hypothesis states that the energy averaged across an ensemble of particles is equal to the energy of a single particle in thermal equilibrium with a heat bath, e.g.\ the detection circuit, averaged over time. The temperature of an eigenmode of a single ion is then defined by the distribution of its repeatedly measured fluctuating energy. If the ion is decoupled from the detection circuit, the energy will be fixed at its instantaneous value, giving enough time for it to be measured.
Repeated measurements of the energy $E$ will follow a Boltzmann distribution
\begin{equation}
p(E) = \frac{1}{k_B T} e^{-E / (k_B T)}\, \text{d}E\,,
\label{eq:boltzman-distribution}
\end{equation}
which can be used to determine the temperature $T_z$ of the axial motion.

Since in Penning traps the frequency is usually independent of the axial energy, the axial temperature can only be measured indirectly: Applying sideband coupling imprints the thermal fluctuations of the axial mode onto the cyclotron mode, whose resulting temperature $T_+$ is then related to the axial temperature by \cref{eq:radial-temperatures}.
Coupling is performed for several seconds to ensure full thermalization of the cyclotron mode.
The cyclotron energy is measured by adiabatically transporting the ion into a second Penning trap with a strong quadratic magnetic field inhomogeneity $B_2 = 43080(120)\,\text{T}/\text{m}^2$. This translates the cyclotron energy into an axial frequency shift
\begin{equation}
\Delta \nu_z = \frac{1}{4 \pi^2 m \, \nu_z} \frac{B_2}{B_0} E_+
\end{equation}
on the order of 10 to 100\,Hz which can be easily measured.
Afterward, the ion is transported back into the initial trap and the cycle is repeated.
A total of 126 measurements were performed for each of three different feedback configurations.
The temperature of the axial mode was then determined via maximum likelihood estimation using the underlying Boltzmann distribution from \cref{eq:boltzman-distribution}.

\begin{figure}
	\includegraphics{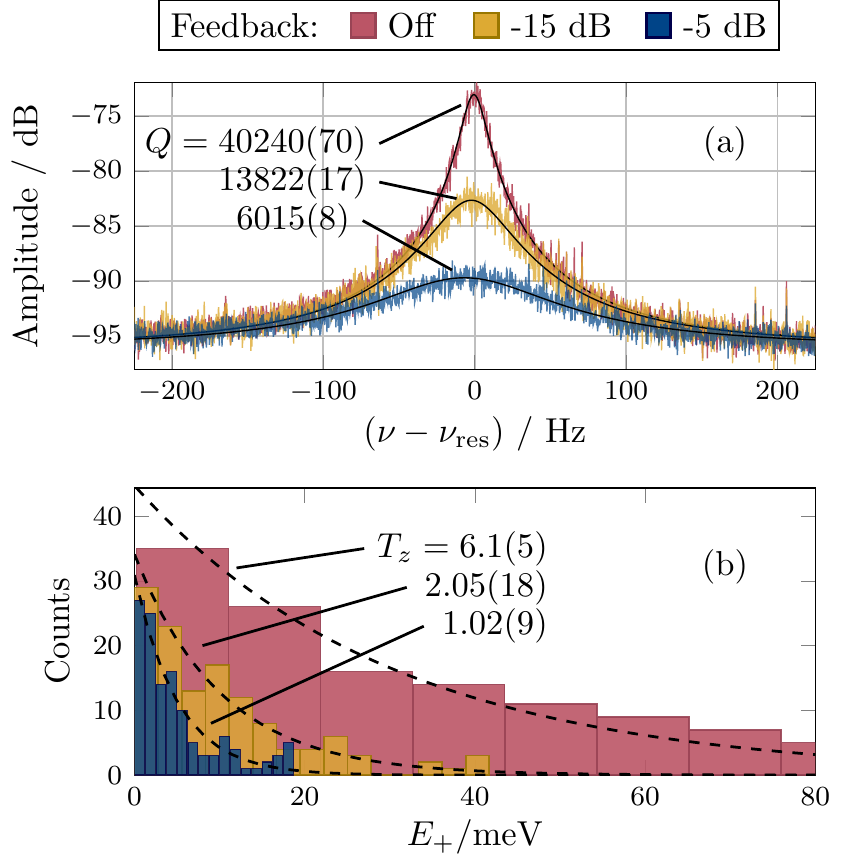}
	\caption{(a) Frequency spectra showing the resonator lineshape for three different feedback configurations. The $Q$ values are determined by curve fits, shown by the black lines. (b) Histogram of measured cyclotron energies for the same feedback gains. The dashed lines show the underlying Boltzmann distribution fitted to the raw data using a maximum likelihood estimation. The resulting temperatures of the axial mode are indicated. The displayed histograms are for visualization purposes only and the binning has no influence on the results.}
	\label{fig:temperature-plots}
\end{figure}

The feedback was configured for a phase shift of $180^\circ$ to lower the $Q$ value of the detection system and thus to cool the ion. The lowered $Q$ value is visible in the spectrum of the resonator shown in \cref{fig:temperature-plots}a.
Applying feedback with a gain of $-15\dB$ cools the ion's axial motion down to $2.05(18)\K$, which corresponds to a reduction by a factor of $3.0(4)$ compared to the measured temperature of $6.1(5)\K$ without feedback. 
This is consistent with the $Q$ value reduction factor of $2.909(6)$, as predicted by \cref{eq:resonator-temperature}. With a feedback gain of $-5\dB$, an axial temperature of $1.02(9)\K$ was measured, corresponding to a cooling factor of $6.0(8)$.
This is slightly less than the $Q$ value reduction factor of $6.686(14)$, which can be attributed to increased feedback of noise, predominantly from the cryogenic amplifier, heating up the ion.

\subsection{Phase Sensitive Measurement of Axial Frequency}

At \pentatrap, the axial frequency is currently measured by fitting a mathematical model of an ion in thermal equilibrium with the detection system to the dip spectrum. As the dip detection technique relies on the incoherent thermal noise of the resonator, its \gls{snr} only increases with $\sqrt t$ over the measurement time $t$, making it a very slow process. A typical axial measurement cycle takes about 7\,min and yields a statistical frequency uncertainty of $10$ to $20\mHz$.
The systematic accuracy depends on the explicit choice of the used mathematical model.

\begin{figure}
	\includegraphics{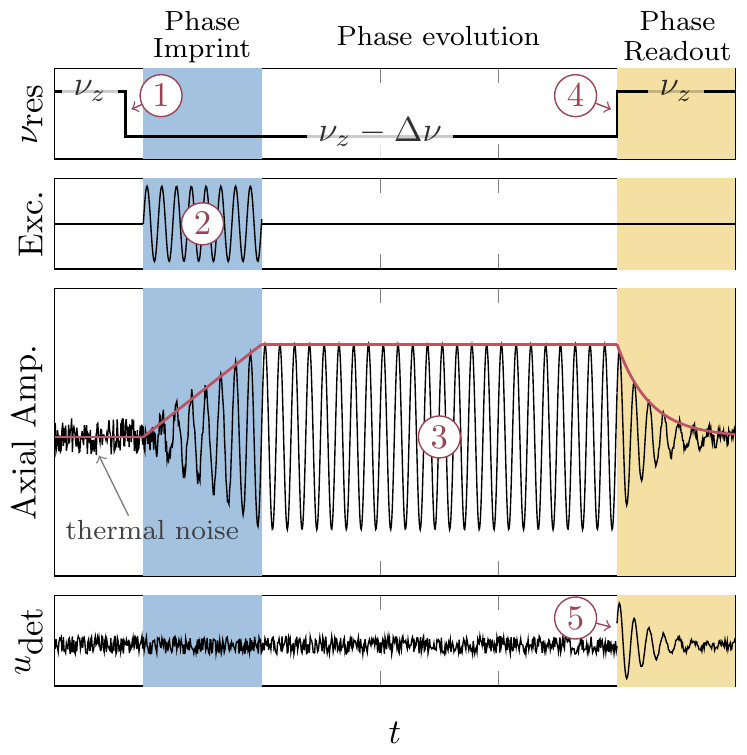}
	\caption{Sketch of the \gls{apnp} measurement scheme (simulated data). The axial ion mode is initially in thermal equilibrium with the detection system, represented by the thermal noise in the axial amplitude, as depicted in the upper \graphpanel. (1) Detuning the center frequency $\nu_\text{res}$ of the resonator away from $\nu_z$ via active feedback leads to a decoupling of the ion from the detection circuit. (2) A \gls{rf} pulse excitation imprints a known starting phase to the axial oscillation. (3) The axial oscillation is allowed to evolve freely for a fixed duration $t_\text{evol}$. Due to the resonator shift, no resistive cooling is occurring and the amplitude stays constant, theoretically allowing unlimited phase evolution times. (4) Switching off feedback brings the detection system back into resonance with the axial oscillation, leading to a measurable signal on its output (5).}
	\label{fig:apnp-scheme}
\end{figure}

An alternative measurement scheme was proposed by \citeauthor{stahlPhasesensitiveMeasurementTrapped2005} [\onlinecite{stahlPhasesensitiveMeasurementTrapped2005}], relying on direct phase measurements of the coherent ion oscillation. As it measures the frequency without the explicit need for a mathematical model, it does not suffer from model-inherited systematic uncertainties. Due to its similarities with the \gls{pnp} technique for cyclotron frequency determination\cite{cornellSingleionCyclotronResonance1989}, it is subsequently referred to as \gls{apnp}. A detailed sketch of this technique is shown in \cref{fig:apnp-scheme}. An \gls{rf} pulse is used to excite the ion and imprint a known starting phase to its oscillation. After a precisely known time $t_\text{evol}$, the phase of the freely evolving axial oscillation is read out, which can be used to determine the axial frequency
\begin{equation}
\nu_z = \frac{2 \pi N + \Delta \varphi}{2 \pi \, t_\text{evol}}\,,
\end{equation}
with $\Delta \varphi$ being the difference between read out phase and imprinted phase and $N$ being the number of whole oscillation cycles within the phase evolution time. $N$ cannot be determined from a single \gls{apnp} cycle, but can be inferred using multiple cycles with varying phase evolution times.
The obtainable frequency precision is given by
\begin{equation}
\label{eq:apnp-sigma-nuz}
\sigma_{\nu_z} = \frac{\sigma_{\varphi}}{2 \pi t_\text{evol}}
\end{equation}
and exceeds the Fourier limit, provided that the phase can be read out with an uncertainty $\sigma_\varphi < 2 \pi$.
The phase evolution time is naturally limited to a few $100\ms$ by the decaying oscillation amplitude due to resistive cooling.
To overcome this limit, active feedback is used to detune the resonator and thus to decouple the ion from the detection system for the duration of phase evolution.
The \gls{apnp} technique was realized using the digital feedback system by connecting its trigger input to a pulse generator, which is already used as the measurement sequencer at \pentatrap, to enable or disable the active feedback with the required high temporal precision.

It was successfully tested with a $^{187}\text{Re}^{32+}$ ion and phase evolution times from $82\ms$ to $2.4\s$. For each evolution time, a set of 20 cycles was acquired to gather information about the statistical phase uncertainty. The results are listed in \cref{tab:apnp-results}. The longest evolution time of $2.4\s$ yields an uncertainty of $20.1\degree$, resulting in an axial frequency uncertainty of $23.3\mHz$, according to \cref{eq:apnp-sigma-nuz}. 
While this is in agreement with the uncertainty of the dip technique, the measurement time is reduced by a factor of 166.
The results show a steady increase of phase uncertainty with longer evolution times, which can be attributed to fluctuations in the trapping potential. 
\begin{table}
	\caption{Statistical uncertainty of the read-out phase for different evolution times $t_\text{evol}$, each acquired with 20 \gls{apnp} cycles.}
	\label{tab:apnp-results}
	\begin{tabular}{lcccccc}
		\toprule
		$t_\text{evol} / \ms$ &$82$ &$205$&$600$&$1100$&$1850$&$2400$ \\
		\midrule
		$\sigma_\varphi$ & $7.4\degree$ & $6.8\degree$ & $7.7\degree$ & $13.0\degree$ & $18.9\degree$ & $20.1\degree$ \\
		\bottomrule
	\end{tabular}
\end{table}
A disadvantage of this technique resides in the relatively high axial oscillation amplitude arising from phase imprinting, which can potentially lead to systematic frequency shifts, for example, due to trap anharmonicities.
This can be compensated for by shrinking the phase space of the initial thermal fluctuations via feedback cooling, which then permits lower excitation amplitudes. This requires the feedback system to be switched between ``cooling configuration'' and ``resonator shift configuration'' with precise timing and without dead time, which is a perfect use-case for the presented digital feedback system. Characterization measurements on this improved technique are pending.

\subsection{Self-Excited Oscillator (SEO)}
\label{sec:seo}

Due to the direct coupling of \gls{fpga} and CPU, the presented digital system is ideally suited for the implementation of a single-ion \gls{seo}. The software control loop to keep the ion amplitude constant is described in \cref{sec:software}. 

The \gls{seo} implementation relies on a complete cancellation of the ion damping, i.e.\ $\gamma \rightarrow 0$, which is accomplished by applying feedback directly to the ion  according to \cref{eq:ion-damping-feedback}. In order to prevent any modifications of the ion detection system the feedback signal must not couple to the resonator. For that, the second output of the feedback system is connected to a separate trap electrode. By carefully tuning the relative phase and gain relation of both channels, the feedback signal, as seen by the ion detection system, can be completely canceled out, while still applying feedback to the ion. 

\begin{figure}
	\includegraphics{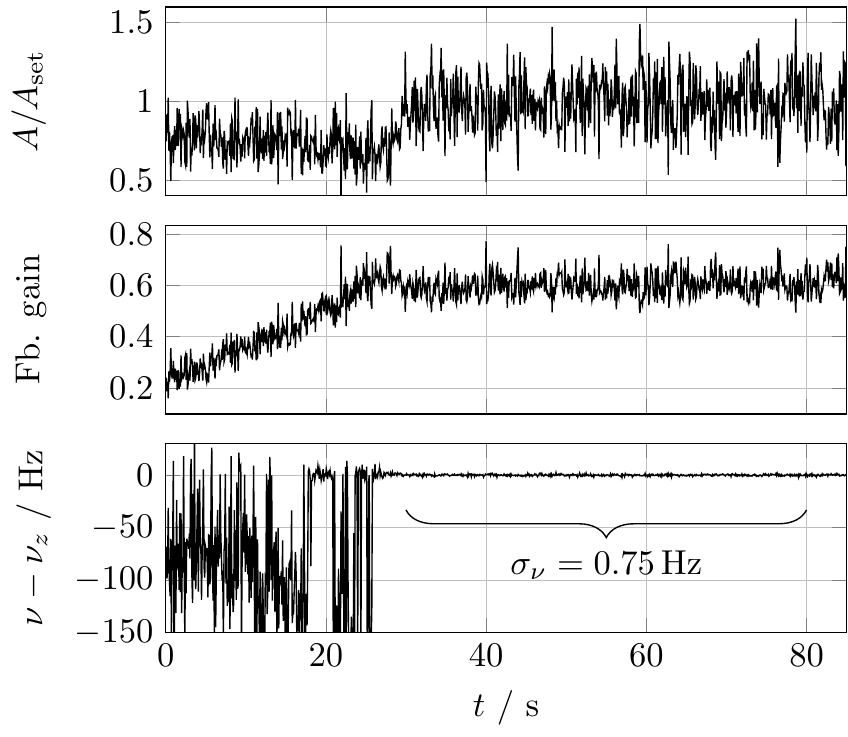}
	\caption{Recording of the startup phase and operation of a single-ion SEO. The first \graphpanel shows the ratio of measured ion amplitude $A$ to the amplitude setpoint $A_\text{set}$. The middle \graphpanel shows the applied feedback gain, which is controlled by a PID loop. The gain is slowly ramped up until the amplitude setpoint is reached at $t\approx 29\s$ and is constantly adjusted afterwards to maintain the desired amplitude. The lowest \graphpanel shows the ion frequency, which is determined from the position of the ion peak in the frequency spectrum.}
	\label{fig:seo}
\end{figure}

The following results were acquired at the \pentatrap experiment using a $^{187}\text{Re}^{32+}$ ion. The amplitude setpoint, at which the ion should be kept constant using the control loop, is set to $3\dB$ above resonator noise level. The internal acquisition system, used for ion amplitude measurement and frequency determination, was configured for a downsampling rate of 4096 and a sample size of 2048, resulting in a control loop interval of $\approx 80\ms$.
The time evolution of the \gls{seo} in operation is shown in \cref{fig:seo}. After the control loop has excited the ion to the desired amplitude, it is held constant for an arbitrary time. The ion now performs coherent oscillation at an amplitude that is easily detectable, enabling very fast axial frequency measurements. While the high ion amplitude comes at the cost of increased systematic shifts, and is thus not suitable for high-accuracy frequency measurements, it can be used to track relative frequency changes at unparalleled measurement rates. This can be used for the detection of spin states\cite{dursoSingleParticleSelfExcitedOscillator2005} or the indirect characterization of trap potential stability by measuring axial frequency fluctuations.

\section{Summary}

Active feedback has proven to be a versatile tool in Penning-trap experiments.
The presented feedback system demonstrates the use of real-time data processing algorithms to manipulate the motion of a single ion confined in a Penning trap. The novel concept utilizes high-speed analog/digital data conversion and a modern \gls{soc} for signal processing, offering high temporal and environmental stability that is only limited by the stability of the used reference frequency and provides surpassing flexibility compared to previously realized analog feedback systems.
By applying feedback to the resonator used within the ion detection system, a $Q$ value variation by two orders of magnitude and a resonance frequency shift of up to $2.78\kHz$ were successfully demonstrated.
Furthermore, it was used for feedback cooling of an ion by a factor of $\approx 6$ down to $1.01(9)\K$.
A full featured acquisition system was implemented inside the \gls{fpga} logic, which transfers the detected signal into the main memory of the attached CPU. Using this feature, the feedback system was successfully used to realize a single-ion self-excited oscillator without the need for auxiliary devices.
The digital architecture allows the feedback parameters to be dynamically varied with precise timing and in synchronization with an external trigger signal, opening up the possibility for new measurement schemes.
The system will be commissioned at \pentatrap, among other things, to implement the presented \gls{apnp} measurement scheme, which was demonstrated to reduce the axial frequency measurement time by two orders of magnitude while eliminating model-related systematics.
As the statistical uncertainty of \pentatrap is currently mainly limited by the axial precision, the presented system will contribute to achieving the mass accuracy goal of a few parts in $10^{12}$.

\begin{acknowledgments}
	We thank our colleagues from the \alphatrap experiment\cite{sturmALPHATRAPExperiment2019}, led by Sven Sturm, for their support and time to carry out the temperature measurements at their experiment.
	
	This work is part of and funded by the Max Planck Society, the Deutsche Forschungsgemeinschaft (DFG, German Research Foundation) - Project-ID 273811115 - SFB 1225, and the DFG Research UNIT FOR 2202. Furthermore this project has received funding from the European Research Council (ERC) under the European Union’s Horizon 2020 research and innovation programme under grant agreement No. 832848 - FunI and we acknowledge funding and support by the International Max Planck Research School for Precision Tests of Fundamental Symmetries (IMPRS-PTFS) and by the Max Planck PTB RIKEN Center for Time, Constants and Fundamental Symmetries.
\end{acknowledgments}

\section*{Data Availability}
The data that support the findings of this study are available from the corresponding author upon reasonable request.

\bibliography{literature.bib}

\end{document}